\documentclass[preprint,preprintnumbers,amssymb,pre,superscriptaddress]{revtex4}
\usepackage{graphicx}
\usepackage{dcolumn}
\usepackage{bm}
\usepackage{amstext}

\newcommand{\trp}[1]{{#1}} 
\newcommand{\hcf}[1]{{#1}} 
\newcommand{\im}{\mathrm{i}}

\begin{document}


\title{Swimming speeds of filaments in nonlinearly viscoelastic fluids}
\author{Henry C. Fu}
\email{Henry_Fu@brown.edu}
\affiliation{Division of Engineering, Brown University, Providence, RI 02912}
\author{Charles W.Wolgemuth}%
\email{cwolgemuth@uchc.edu}
\affiliation{Department of Cell Biology, University of Connecticut Health Center, Farmington, CT 06030}%
\author{Thomas R. Powers}
\email{Thomas_Powers@brown.edu}
\affiliation{Division of Engineering, Brown University, Providence, RI 02912}


\date{June 12, 2008}


\begin{abstract}
Many microorganisms swim through gels and non-Newtonian fluids in
their natural environments.  In this paper, we focus on microorganisms
which use flagella for propulsion.  We
address how swimming velocities are affected in nonlinearly
viscoelastic fluids by examining the problem of an infinitely
long cylinder with arbitrary beating motion in the Oldroyd-B
fluid.  We solve for the 
swimming \trp{velocity} 
in the limit in
which deflections of the cylinder from its straight configuration are small
relative to the radius of the cylinder and the wavelength of the
deflections; 
\trp{furthermore,} the radius of the cylinder is small
compared to the wavelength of deflections.   We find that swimming
velocities are diminished by nonlinear viscoelastic effects.  We apply
these results to examine what types of swimming motions can produce net
translation in a nonlinear fluid, comparing to the Newtonian case, for
which Purcell's ``scallop'' theorem describes how time-reversibility
constrains which swimming motions are effective.  We find that the
leading order violation of the scallop theorem occurs for reciprocal
motions in which the backward and forward strokes occur at different
rates.    
\end{abstract}


\maketitle

\section{Introduction}

\trp{Bacteria and other microscopic swimmers live in a world of low Reynolds number, in which viscous effects dominate over inertia~\cite{Ludwig1930,purcell1977}. In this regime, the drag forces acting on a swimmer are much greater than the forces necessary to achieve the observed accelerations of these swimmers. At the limit of zero Reynolds number, the sum of the net drag force and any other external forces, such as buoyancy, must vanish.  Therefore, the physics underlying the swimming strategies of microorganisms is markedly different from that of large organisms such as fish. For example, in the absence of inertia, the diffusion of viscous shear waves is essentially instantaneous. Thus, the} 
flows induced by a swimming creature's
motions start and stop immediately when the creature's movements start
and stop.  
\trp{This phenomenon is closely related to the} 
 ``scallop theorem," 
 of Purcell
\cite{purcell1977},
which states that time reversible strokes \trp{in a Newtonian fluid, such as water,} cannot produce net swimming~\cite{Ludwig1930}, because there is an equal and
opposite net translation for corresponding forward and backward stroke
motions.  As a result, \trp{swimming} microorganisms 
use \trp{strokes} 
with a clear time direction, \trp{such as the traveling wave patterns on the flexible flagella of mammalian sperm, or the rigid-body rotation of the helical flagellar filaments of  \textit{Escherichia coli}.}

\trp{Early theoretical work elucidated the basic principles of
  swimming at zero Reynolds numbers in a Newtonian fluid. For example,
  Taylor calculated the swimming speed of an infinite two-dimensional
  sheet as a function of the amplitude, wave number, and wave speed of
  a traveling wave on the sheet~\cite{taylor1951}. He also did the
  same calculation for an infinite cylindrical
  filament~\cite{gi_taylor1952}. These calculations showed that the
  swimming speed is proportional to the wave speed, as expected from
  dimensional analysis. Since the shape and speed of the waves on the sheet or filament in these calculations are independent of the load, the swimming speed is independent of the viscosity of the fluid. In other work, various groups have calculated how viscous forces and forces internal to a flagellum determine the shape of flexible filaments actuated at one end~\cite{machin1958,wiggins_goldstein1998,YuLaugaHosoi2006} or actuated by internal motors distributed along the filament~\cite{camalet_et_al1999,camalet-julicher00,Riedel-Kruse2007,FuWolgemuthPowers2008Shapes}.}

\trp{All the calculations just mentioned apply to Newtonian fluids. However, swimming microorganisms often encounter elastic fluids.} For example, mammalian sperm must swim through 
viscoelastic mucus in the female reproductive tract~\cite{HoSuarez2003,
  Suarez_et_al1991, SuarezDai1992}; the
bacterium \textit{Helicobacter pylori} must swim through 
mucus
lining the stomach~\cite{MontecuccoRappuoli2001}; and spirochetes must burrow through connective
tissue in the course of infection~\cite{WolgemuthCharonGoldstein2006}.  \trp{The theory of swimming in non-Newtonian fluids is much less developed than the Newtonian case. }
\trp{Fulford, Katz, and Powell extended the resistive force theory of Gray and Hancock~\cite{gray_hancock1955} to the case of a linear Maxwell model for viscoelasticity, and found no change in the swimming speed compared to the Newtonian case~\cite{FulfordKatzPowell1998}. More recently, Lauga emphasized that since swimming speed depends nonlinearly on wave amplitude, a realistic treatment of swimming in a non-Newtonian medium should include a nonlinear model for the viscoelastic medium~\cite{Lauga2007}. Lauga investigated several nonlinear fading memory models and found that for a given wave pattern on a two-dimensional sheet, viscoelastic effects slow the swimmer relative to its speed in a Newtonian liquid~\cite{Lauga2007}.}


\trp{In our previous work, we calculated the swimming velocity of an
  infinite filament with a prescribed waveform in the Upper Convected
  Maxwell Model~\cite{BirdArmstrongHassager1987}, a simple nonlinear
  model for a viscoelastic material with fading
  memory~\cite{FuPowersWolgemuth2007}. We also studied the effect of
  the viscoelastic medium on the shape of a beating filament described
  by a simple sliding filament model~\cite{camalet_et_al1999}, and
  showed how altered beating shapes affect swimming velocities. In the current paper, we give a more detailed account of our calculation of the swimming speed of an infinite filament with a prescribed waveform.  We refine our previous calculation by including the effects of solvent viscosity with the Oldroyd-B model for a nonlinearly viscoelastic polymer solution. }
\trp{As an application of our calculations, we examine how the scallop
  theorem is spoiled in a nonlinearly viscoelastic fluid at zero
  Reynolds number.  In this paper we consider prescribed shapes of
  beating patterns.  In
  actual swimmers with beating filaments, the shape is determined by
  the interaction of filament structure with hydrodynamic forces,
  which we address in another publication~\cite{FuWolgemuthPowers2008Shapes}.
  Here we take the point of view that the beating pattern is already
  known (for example, from direct observation) and ask what the
  consequences are for swimming properties.  We restrict our
  discussion to time-periodic motions and associated flows,
  corresponding to repetitive swimming actions that can be employed to
  move an arbitrarily large distance by a self-contained organism.
  For these time-periodic motions, the associated flows will have both
  time-periodic and constant
  components.  Only the constant, time-averaged component
  produces net translation over many cycles.}

\trp{We close this introduction with a summary of the structure of
  this paper. In Sec.~\ref{models} we review the Oldroyd-B model and
  how it naturally arises for a material with fading memory. Then, in
  the next section, we calculate the flow fields induced by a helical
  traveling wave on an infinite cylinder. We work to second order in
  the deflection of the cylinder, and deduce the time-averaged
  swimming speed. In Sec.~\ref{ScallopThm} we explore the
  ramifications of our calculation for the scallop theorem, and show
  quantitatively how a reversible stroke pattern leads to
  propulsion. Section~\ref{discuss} is the conclusion. Finally, in the
  appendix we calculate the viscoelastic force per unit length to
  first order acting on a filament with transverse or longitudinal
  (compressional) traveling waves. }


\section{Nonlinearly viscoelastic fluids}
\label{models}
\trp{Since we consider the zero Reynolds number regime, inertia is irrelevant and the dynamics of an} incompressible
medium is governed by force balance, 
$-\nabla p +\nabla\cdot{\bm \tau} =0 $, where $p$ is the pressure
associated with incompressibility, $\nabla \cdot \mathbf{v} = 0$; $\mathbf{v}$ is the flow velocity;
and $\bm \tau$ is the deviatoric stress tensor. \trp{The particular
  character of a medium is determined by the constitutive relation,
  which relates the stress to the strain and rates-of-strain. In this paper we consider
  dilute polymer solutions, in which the stress relaxes to the stress
  of a Newtonian viscous fluid over a time scale $\lambda$. A simple constitutive relation for a dilute polymer solution is the single-relaxation-time Maxwell model:}
\begin{equation}
\tau_{ij}+\lambda\frac{\partial\tau_{ij}}{\partial t}=\eta\dot\gamma_{ij},\label{MM}
\end{equation}
\trp{where $i$ and $j$ label Cartesian coordinates, $\eta$ is
  viscosity, and $\dot\gamma_{ij}=\partial v_i/\partial x_j+\partial
  v_j/\partial x_i$ is the strain rate. This constitutive relation is
  valid to first order in deformation.  However, the swimming velocity
  of a prescribed motion is second order in the
  deformation~\cite{taylor1951,gi_taylor1952}. To see why, consider a
  traveling wave of amplitude $h$ on an infinite cylindrical filament
  (Fig.~\ref{taylorcylinder}). The swimming velocity may be
  represented as a power series in $h$. Replacing $h$ with $-h$
  changes the sign of the first order contribution to the velocity. On
  the other hand, changing the sign of $h$ amounts to translating the
  filament by half of a wavelength, and therefore yields no change in
  the time-averaged swimming velocity. Thus we conclude that the
  swimming velocity is quadratic in $h$ to leading order, \hcf{ and
  therefore nonlinear corrections to Eq.~(\ref{MM}) will be important 
for the study of swimming in a viscoelastic medium~\cite{Lauga2007}.}}

\trp{It is useful to recall why the constitutive relation~(\ref{MM}) is inapplicable for large deformations. First, note that the Maxwell constitutive relation does not respect Galilean invariance. This defect is easily fixed by replacing the partial time derivative of the stress, $\partial\tau_{ij}/\partial t$, with the material derivative, $\partial\tau_{ij}/\partial t+\mathbf{v}\cdot\nabla\tau_{ij}$. However, this modified constitutive relation is still unsuitable for working beyond first order in deformation because it is sensitive to the overall orientation of the material. For example, consider a steady-state flow, for which (\ref{MM}) predicts that $\tau_{ij}=\eta\dot\gamma_{ij}$. Now place the medium on a slowly rotating turntable~\cite{BirdArmstrongHassager1987}. Since viscous forces are much greater than the fictitious forces arising from rotation, the relation between stress and strain rate in the rotating material is the same as that in the stationary material, to an excellent approximation~\cite{DeGennes1983}: $\tau'_{ij}=\eta\dot\gamma'_{ij}$, where the prime denotes the quantities for the rotating material. On the other hand, the components $\tau'_{ij}$ are time-dependent, because $\tau'_{ij}=R_{ik}R_{jl}\tau_{kl}$, where $R_{ij}$ is a time-dependent rotation matrix. Therefore, Eq.~(\ref{MM}) cannot hold for the primed quantities, even with the partial time derivative replaced with the material derivative.}

\trp{In general, changes in the components $\tau_{ij}$ due to motion
  of the basis vectors should not enter the constitutive relation. The
  formulation of rheological equations of state requires a time
  derivative that does not depend on the choice of
  basis~\cite{oldroyd1950}. Any constitutive relation that reflects
  the small magnitudes of fictitious forces must be formulated in
  terms of time derivatives of stress or strain that transform
  homogeneously under change of coordinates.  \hcf{ There are many different
  ways to construct a time derivative of the stress tensor that is
  insensitive to the rate of change of the basis vectors.  In the
  following we present one way to do so which leads to the
  nonlinear viscoelastic model used in the rest of the paper.}
}

\trp{
\hcf{
Due to the small magnitudes of fictitious forces, the constitutive
relation should be insensitive not only to global rotations such as
that of the turntable described above, but also to local rotations and
translations of fluid elements.
One way to enforce this more general condition is to express the time
derivatives in terms of coordinates which track fluid elements.} 
Suppose $\{\xi^1,\xi^2,\xi^3\}$ are convective coordinates, with
$\mathbf{r}(\xi^1,\xi^2,\xi^3,t)$ the position at time $t$ of the
particle with Cartesian coordinates $\{\xi^1,\xi^2,\xi^3\}$ at the
initial time.  Convective coordinates are also known as Lagrangian coordinates. The basis vectors associated with convective coordinates are 
$\mathbf{g}_i=\partial\mathbf{r}/\partial\xi^i$, and the stress takes the form $\bm{\tau}=\tau^{ij}\mathbf{g}_i\otimes\mathbf{g}_j$, or more simply $\bm{\tau}=\mathbf{g}_i\tau^{ij}\mathbf{g}_j$. For Cartesian coordinates, whether an index is raised or lowered has no significance; however, for general coordinates, raised indices signify contravariant coordinates, and lowered indices signify covariant coordinates~\cite{BirdArmstrongHassager1987}.
Since $\partial \mathbf{g}_i/\partial t=\partial\mathbf{v}/\partial\xi^i=(\mathbf{g}_i\cdot\nabla)\mathbf{v}$, we have
\begin{equation}
\left.\frac{\partial\bm\tau}{\partial t}\right|_{\xi^i}=\mathbf{g}_i\left(\frac{\partial\tau^{ij}}{\partial t}\right)\mathbf{g}_j+\left(\mathbf{g}_i\cdot\nabla\mathbf{v}\right)\tau^{ij}\mathbf{g}_j
+\mathbf{g}_i\tau^{ij}\left(\mathbf{g}_j\cdot\nabla\mathbf{v}\right).
\label{dtaudt}
\end{equation}
The second two terms in Eq.~(\ref{dtaudt}) arise from the the dependence of the basis vectors on time and therefore they cannot enter the constitutive relation. To eliminate the spurious time-dependence, define the upper-convected time derivative using convective coordinates:
\begin{equation}
\stackrel{\triangledown}{\bm\tau}\equiv\left.\mathbf{g}_i\frac{\partial\tau^{ij}}{\partial t}\right|_{\xi^i}\mathbf{g}_j.\label{uc-convcoords}
\end{equation}
In coordinate-free form, the upper-convected derivative is
\begin{equation}
\stackrel{\triangledown}{\bm\tau}
=\frac{\partial \bm{\tau}}{\partial t}
+\mathbf{v}\cdot\nabla{\bm\tau}
-{\bm\tau}\cdot\nabla\mathbf{v}
-(\nabla\mathbf{v})^{\mathrm{T}}\cdot{\bm\tau}.\label{u-convected}
\end{equation}
It follows directly from Eq.~(\ref{uc-convcoords}) that the components
of $\stackrel{\triangledown}{\bm\tau}$ in a general coordinate system
transform homogeneously \hcf{even under a time-dependent} change of coordinates.
}

\trp{The upper-convected time derivative is just one way of many to construct a time derivative of the stress tensor that is insensitive to the rate of change of the basis vectors. For example, if we consider the covariant components $\tau_{ij}$ of the stress tensor in the convected-coordinate basis, a different natural derivative emerges: the lower-convected derivative~\cite{Morrison2001}. The choice of one particular convective derivative over another cannot be justified on symmetry grounds alone, but must be made using information from experiments or the continuum limit of a microscopic model.
} 

\trp{For example, a commonly used model for dilute polymer solutions
  is the Oldroyd-B model~\cite{BirdArmstrongHassager1987}. We will use
  this model in this paper. In the Oldroyd-B model, the polymer
  component is represented by the Upper Convected Maxwell model, which
  is the Maxwell constitutive equation (\ref{MM}) with the time
  derivative replaced by the upper-convected time derivative.  The
  appearance of the upper-convected time derivative, rather than some
  other form insensitive to rates of change of basis vectors, can be
  justified by appealing to a microscopic model of polymer stresses
  arising from Hookean dumbbells~\cite{BirdArmstrongHassager1987b}.  The solvent is represented by the usual Newtonian constitutive relation for a viscous fluid. Thus, the deviatoric stress for the solution is }
\begin{equation}
{\bm \tau} + \lambda \stackrel{\triangledown}{\bm  \tau} = \eta \dot {\bm \gamma}
+ \lambda \eta_\mathrm{s} \stackrel{\triangledown}{\dot {\bm \gamma}}, \label{OldroydB}
\end{equation}
where the total viscosity $\eta = \eta_\mathrm{p} +
\eta_\mathrm{s}$ is the sum of polymer and solvent viscosities.  
\hcf{Note that if a patch of fluid undergoes a spatially uniform
  rotation or translation, due to the covariant nature of this
  equation, within the patch the constitutive relation is insensitive
  to the motion.   
In (\ref{OldroydB}),}  
if the solvent viscosity is ignored ($\eta_s = 0$
), the Oldroyd-B fluid reduces to the Upper Convected Maxwell fluid, and if
there is no polymer ($\eta = \eta_s$) the Oldroyd-B fluid reduces to a
Newtonian fluid.  
The model for the Oldroyd-B fluid 
\trp{predicts} realistic elastic
and first normal stress effects as long as elongational flows are
not large; however, it does not capture shear-thinning or yield-stress
behaviors observed in some non-Newtonian fluids.  The Oldroyd-B model
has been used to describe Boger fluids \cite{BinningtonBoger1986}.      
We use the Oldroyd-B fluid as an example.  The choice of constitutive
  relation ultimately depends on accurately modeling the properties of
  particular real-life fluids and flows.  We note that in the case of
  an infinite sheet, Lauga examined a number of nonlinear 
  models, including Finite Extensible Nonlinear Elastic models which
  remedy the limitations of Oldroyd-B in elongational flows, and found that the swimming speed was not affected by the
  choice of constitutive model~\cite{Lauga2007}.

\section{Swimming velocity of filaments in viscoelastic fluids}

In 
\trp{this section} we describe the solution to the problem of an
infinitely long cylinder with prescribed beating motion in a
viscoelastic fluid.  We obtain the swimming velocity in a nonlinearly
viscoelastic fluid, and find that in general it is decreased relative
to the swimming velocity 
\trp{in} a Newtonian fluid.  \trp{Our results are different from} 
the results of Fulford, Katz, and Powell \cite{FulfordKatzPowell1998},
who 
\trp{used the constitutive relation~(\ref{MM}) to find that a filament with prescribed beating pattern swims at the same speed in viscous and linearly viscoelastic fluids.}
\trp{Our results are} in \trp{broad} agreement with \trp{those} 
of Lauga,
\cite{Lauga2007} who 
\trp{solved} the corresponding problem of a\trp{n infinite} sheet with 
\trp{traveling-wave} 
displacements. 

 \trp{Figure~(\ref{taylorcylinder}) shows a snapshot of the swimming filament. The beating motion is represented by a transverse traveling wave. This motion induces flow. To calculate the swimming velocity, we work in the frame of the swimmer and calculate the $z$-component of the flow far from the swimmer. As argued above, the swimming velocity is second order in the amplitude of the deflections of the filament; therefore we must calculate the $z$-component of the flow to second order. We calculate the average speed over one period of the motion of the filament.} 

\subsection{First order solutions for a cylinder with prescribed beating pattern}\label{cylinderforces}

\begin{figure}
\includegraphics[width = 3.5in]{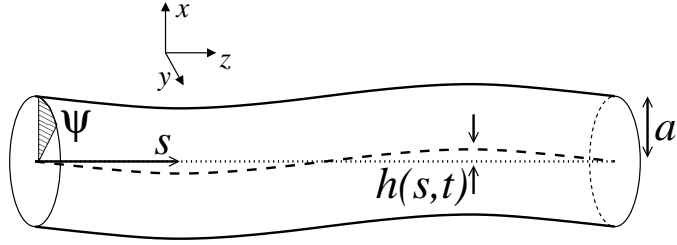}
\caption{
\trp{A section of an infinite filament} with prescribed beating pattern $h(s,t)$.
} \label{taylorcylinder}
\end{figure}

\trp{In this subsection we define the deformation of the filament, prescribe the boundary conditions, and solve for the flow velocity to first order in the deformation of the filament.}
We consider a cylinder with radius $a$, with its axis lying along the
$\hat \mathbf{z}$ direction.  For the undisturbed, straight cylinder,
material points \trp{on the surface} are parameterized by $s$, the distance along the axis,
and the angle $\psi$ (see 
Fig. \ref{taylorcylinder}). \trp{Note that these coordinates are convective coordinates for the filament surface.}   For a transverse \trp{planar}
traveling wave, 
\trp{the deformation of the}
cylinder is given by prescribing the position $s\hat\mathbf{z}+h(s,t) \hat {\mathbf{x}}$ of the centerline of the
cylinder. 
Material points \trp{on} 
the \trp{surface of the} cylinder are 
\trp{parameterized} by
\begin{eqnarray}
{\bf r}(\psi,s,t) &=&  \hat
  {\bf x}[h(s,t)+ a \cos \psi] +\hat {\bf y} a \sin\psi  + \hat{\bf z}s ; \nonumber \\  
h(s,t) &=& \mathrm{Re} \lbrace \tilde h \exp (\mathrm{i}q s - \mathrm{i}\omega t)\rbrace.\label{transverseh}
\end{eqnarray}
Because the displacement is solely in the $\hat {\bf x}$ direction,
the filament must be extensible. 
\trp{Enforcing} inextensibility leads to motion of the filament
surface in the $\hat{\bf{z}}$ direction with amplitude of order
$h^2$; this type of longitudinal motion would give rise to corrections
to the swimming velocity of order $h^4$, higher order than the results
we report in this paper.   
For technical ease, we solve for arbitrary motion by decomposing into
a linear superposition of traveling waves; therefore it is sufficient
to assume that the prescribed shape change in Eq.~(\ref{transverseh}), 
is given by a single propagating mode.  As we will show, the swimming
velocity, although second order in h, has a form amenable to this
decomposition into traveling waves. 
We also introduce notation for variables with
tildes that we 
use in the remainder of this paper: $A = \mathrm{Re} \lbrace \tilde A \exp{(\mathrm{i}qs-\mathrm{i}\omega t)}
\rbrace$.

\begin{figure}
\includegraphics[width = 1.5in]{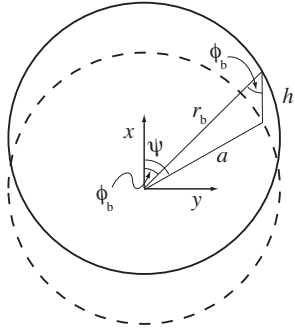}
\caption{
\trp{A cross-section of the infinite filament, showing the undeformed (dashed line) and current configuration (solid line), and the relation between $\phi_\mathrm{b}$ and the convective coordinate $\psi$.} 
} \label{shift}
\end{figure}

Assume that the displacement of the filament from
its undisturbed state is small compared to the smallest wavelength of
the cylinder motion $h(s,t)$  and the cylinder radius $a$.  This \trp{assumption}
allows us to take advantage of the cylindrical symmetry. \trp{Later, we will find it convenient to further demand that $qa\ll1$.} Using cylindrical coordinates, \trp{with basis vectors $\hat{\bf z}$ and
\begin{eqnarray}
\hat{\bf r}&=&\hat{\bf x}\cos\psi+ \hat{\bf y}\sin\psi\\
\hat{\bm \phi}&=&-\hat{\bf x}\sin\psi+\hat{\bf y}\cos\psi,
\end{eqnarray}
denote} the position of \trp{points on} the surface 
by
$(r_\mathrm{b},\phi_\mathrm{b},z_\mathrm{b})$.
\trp{(The subscript $\mathrm{b}$ stands for ``boundary.")}   To first order in $h$ (see Fig.~\ref{shift}), 
\begin{eqnarray}
r_\mathrm{b} &=& a +  h \cos \psi  \nonumber \\
\phi_\mathrm{b} &=& \psi -  \frac{h}{a} \sin \psi  \nonumber \\
z_\mathrm{b} &=& s. \label{boundaryposition}
\end{eqnarray}
\trp{For} 
no-slip boundary conditions on the filament surface,
\begin{equation}
{\bf v}( {\bf r}(\psi,s,t) ) = \left.\frac{\partial\mathbf{r}}{\partial t}\right|_{\psi,s}.\label{BC}
\end{equation}
\trp{Since our goal is to calculate the swimming velocity to leading order, we must expand the boundary condition (\ref{BC}) to second order in the dimensionless displacement $q\tilde h$. To all orders in $q \tilde h$, the right-hand side of Eq.~(\ref{BC}) equals $\dot h\hat{\bf x}$, where here and elsewhere the dot signifies the partial time derivative. Since the leading order contribution to the flow velocity is first order in $\tilde h$, we need only expand the left-hand side of Eq.~(\ref{BC}) to first order in the argument of the flow velocity: 
\begin{equation}
\left[\mathbf{v} +h\hat{\bf x}\cdot\nabla\mathbf{v}\right]_{a,\psi,s,t}=\dot h\hat{\bf x}.
\end{equation}
The flow velocity at a fixed point also has an expansion in powers of $q\tilde h$:
\begin{equation}
\mathbf{v}=\mathbf{v}^{(1)}+\mathbf{v}^{(2)}+ ....
\end{equation}
Thus the boundary condition to first order is
\begin{equation}
\mathbf{v}^{(1)}(a,\psi,s,t)=\dot h\hat{\bf x}.\label{bc1st}
\end{equation}
}
%

To obtain the forces and swimming velocity, we need to solve the 
Stokes 
equations \trp{mentioned above,} $\nabla\cdot{\bm \tau}-\nabla p  =0$ and
$\nabla \cdot \mathbf{v} = 0$,
where $\bm \tau$ is \trp{now} the deviatoric stress tensor \trp{that satisfies}  
Eq.~(\ref{OldroydB}). 
\trp{To simplify the discussion of the equations of motion for the fluid}, we 
nondimensionalize by measuring length in
units of $1/q$, time in units of $1/\omega$, and pressure and
stress in units of $\eta \omega$.  Hereafter in this and the next
section, we use the same variables to
denote the nondimensional quantities, unless explicitly stated
otherwise.  \trp{At linear order, all time-dependent quantities are proportional to $\exp(-\im t)$.}
\trp{To} first order, \trp{Eq.~(\ref{OldroydB})} is 
\begin{equation}
 \tilde {\bm \tau}^{(1)} = \frac{1 -\mathrm{i} \mathrm{De}_2}{1 -
 \mathrm{i} \mathrm{De}} \tilde{\dot{\bm \gamma}}^{(1)},
\end{equation}
where the Deborah numbers $\mathrm{De} = \omega \lambda$ and
$\mathrm{De}_2 = \mathrm{De} \eta_s/\eta$.  
\trp{The first-order flow obeys $\nabla\cdot\mathbf{v}^{(1)}=0$ and}
\begin{eqnarray}
\nabla^2 \left( {\tilde {\mathbf{v}}}^{(1)}  e^{-\im(t -
  z)}\right)-\frac{1-\im \mathrm{De}}{1-\im \mathrm{De}_2}\nabla
\left( \tilde{p}^{(1)} e^{-\im(t -  z)}\right)=0.
\end{eqnarray}
\trp{The first-order boundary condition (\ref{bc1st}) reduces to} $\tilde {\mathbf
  v}^{(1)}(a,\psi) = -\im \tilde h \hat {\mathbf x}$, where ${\mathbf v}^{(1)}=\mathrm{Re}\lbrace \tilde {\mathbf
  v}^{(1)}(r,\phi)\exp(\im z-\im t)\rbrace$ in accord with the convention of Eq.~(\ref{transverseh}). 

The solutions to these equations in cylindrical coordinates are known \cite{gi_taylor1952,
  happel_brenner1965}.  Note that although there is no solution to the Stokes equations for dragging a rigid infinite rod through fluid~\cite{landauFM}, the finite wavelength together with the rod radius provide a cutoff in our problem.
The first order results for the \trp{pressure} and velocity field 
are presented in Table \ref{firstorder}.   In the table, $K_0(r)$ and
$K_1(r)$ are modified Bessel functions of the second kind.  These results are valid up to
leading order in an expansion in $1/ \log(a)$.  The flow fields are the same as
those in the Newtonian case \cite{gi_taylor1952}, but the pressure
has an additional complex factor.  

\begin{table*}
\begin{tabular}{|c||c|}
\hline \\
$
\tilde p $& 
$
-\mathrm{i} \tilde h \cos \phi A  K_1(r) $\\ [1ex]

$
\tilde v_r $& 
$
-\mathrm{i} \tilde h \cos \phi \left[ \alpha A r K_1(r)+BK_2(r)+CK_0(r)\right] $\\
[1ex]
$
\tilde v_\phi $& 
$
-\mathrm{i} \tilde h \sin \phi \left[ BK_2(r)-CK_0(r) \right] $\\[1ex]
$
\tilde v_z $& 
$
-\tilde h \cos \phi \left[ \alpha A rK_0(r) +\left(B+C-\alpha A\right)K_1(r)\right] $ \\[1ex]
$\alpha A$ & $\left\{K_0(a)+aK_1(a)\left[\frac{1}{2}+\frac{K_0(a)}{2K_2(a)}-\frac{K_0(a)^2}{K_1(a)^2}\right]\right\}^{-1}$ 
\\[1ex]
$B$&$-\alpha Aa K_1(a)/\left[2K_2(a)\right]$\\[1ex]
$C$&$\left[1-aK_1(a)\alpha A/2\right]/K_0(a)$\\[1ex]
\hline 
\end{tabular}
\caption{
 \trp{Pressure and velocity field for a cylinder with a transverse traveling wave; $\alpha=(1-\mathrm{i De})/(1-\mathrm{i De}_2)$.}
  \label{firstorder}}
\end{table*}

These first order solutions are \trp{necessary for the calculation of the swimming velocity of the filament, which we present in the next subsection. These solutions are also} sufficient for calculating the lowest
order effects of viscoelasticity on hydrodynamic forces and the
shapes of flexible swimmers~\cite{FuWolgemuthPowers2008Shapes}. 
\trp{In appendix~\ref{forces1st}, we calculate the force per unit length acting on filaments with transverse and longitudinal traveling waves. These results are consistent with the resistive force theory of Fulford, Katz, and Powell~\cite{FulfordKatzPowell1998}.}


\subsection{Velocity of a cylinder with prescribed beating pattern} \label{2ndOrderSection}

\trp{Turning now to the second-order calculation, we expand the boundary condition~(\ref{BC}) to second order:}
\begin{equation}
\left[\mathbf{v}^{(2)}+h\hat\mathbf{x}\cdot\nabla\mathbf{v}^{(1)}\right]_{a,\psi,s,t}=0.\label{BC2}
\end{equation}
To isolate the nonlinear \trp{terms of the equation of motion,} 
it is convenient to rewrite the
convected time derivatives 
as 
\begin{eqnarray}
\stackrel{\triangledown}{\bm \tau} &=& \partial_t {\bm \tau} + \mathbf{T}\\
\stackrel{\triangledown}{\dot {\bm \gamma}} &=& \partial_t {\dot {\bm \gamma}} + \mathbf{G}.
\end{eqnarray}
\begin{widetext}
The second-order 
equations are
\begin{equation}
(1 + \mathrm{De} \, \partial_t) {\bm \tau}^{(2)} =   (1 +
  \mathrm{De}_2 \partial_t) \dot {\bm
  \gamma}^{(2)} - \mathrm{De} {\bf T}^{(2)} + \mathrm{De}_2 {\bf
  G}^{(2)} , 
  \end{equation}
or, invoking force balance $-\nabla p^{(2)}+\nabla\cdot{\bm \tau}^{(2)}=0$,  
  \begin{equation}
{(1 + \mathrm{De} \, \partial_t)} \nabla { p}^{(2)}
 = {(1 + \mathrm{De}_2 \partial_t)}\nabla^2 {\bf v}^{(2)} - {\mathrm{De}} \nabla
  \cdot {\bf T}^{(2)}  + {\mathrm{De}_2} \nabla \cdot {\bf G}^{(2)}. \label{2ndOrderFlow}
\end{equation}
\end{widetext}
Note that ${\bf T}^{(2)}$ and ${\bf G}^{(2)}$ only have contributions from
first order flows and stresses.   

\trp{We must solve Eq.~(\ref{2ndOrderFlow}) for $v_z^{(2)}$.
The time \trp{dependence of $v_z^{(2)}$
is most easily expressed by}
expanding in modes $e^{i m(z-t)}$.   In our first order calculations, 
\trp{only $m=1$ appeared}.  Second order involves $m =
0,\,\pm2$.  Similarly, the azimuthal dependence of velocity and
stress fields can 
be expanded in modes of $\cos(n\phi)$ and
$\sin(n\phi)$, with $n=0$ and $n=\pm2$ appearing at second order.}
To find the swimming velocity, it is only necessary to solve for the
time- and $\phi$-averaged $z$-component of the velocity field, 
which corresponds to the net flow in the $z$-direction \trp{far from the filament}.  This \trp{net flow} is the
$m=0$, $n=0$ mode.   
\trp{In} the $m=n=0$ part of the second-order equation
(\ref{2ndOrderFlow}), all terms with $\partial_{\phi}$, $\partial_t$, and $\partial_z$ drop out,
leaving
\begin{widetext}
\begin{equation}
\frac{1}{r} \partial_r \left[r \partial_r  v_z^{(2)(m=n=0)} \right] =\mathrm{De}
     \frac{1}{r} \partial_r  \left[  r T_{rz}^{(2)(m=n=0)}- \frac{\mathrm{De}_2}{\mathrm{De}} r G_{rz}^{(2)(m=n=0)}\right],  \label{aveq}
\end{equation}
\trp{where $T_{rz}$ can be expressed in cylindrical coordinates as~\cite{BirdArmstrongHassager1987}}
\begin{equation}
T_{rz} = v_r \partial_r \tau_{zr} + v_z \partial_z \tau_{zr} +
\frac{v_{\phi}}{r} \partial_{\phi} \tau_{zr} - \tau_{rr} \partial_r
v_z - \tau_{zr} (\partial_z v_z + \partial_r v_r) - \tau_{zz}
\partial_z v_r - \frac{\tau_{r \phi}}{r} \partial_{\phi} v_z -
\frac{\tau_{\phi z}}{r} \partial_{\phi} v_r. 
\end{equation}
\end{widetext}
\trp{A similar expression holds for $G_{rz}$, with the components of ${\bm\tau}$ replaced by the corresponding components of $\dot{\bm\gamma}$. Examination of the real part of $G_{rz}$ reveals that every term is proportional to an odd power of $\sin(z-t)$ or $\cos(z-t)$; therefore, the average of $G_{rz}$ over a period vanishes.
Integrating Eq.~(\ref{aveq}) over $r$ yields}
\begin{equation}
v_z^{(2)(m=n=0)}  = \mathrm{De} \int \! {\mathrm d}r \,
T_{rz}^{(2)(m=n=0)}  
-U,
\end{equation}
\trp{where $U$ is a constant.}
\trp{The complete expression for $T_{rz}^{(2)}$ in terms of $a$ and the modified Bessel functions $K_0(r)$ and $K_1(r)$ may be constructed from the entries of Table~\ref{firstorder}. To simplify the resulting expression, it is convenient to expand the coefficients $\alpha A$, $B$, and $C$ in powers of $a$. In this expansion, we only keep the leading logarithms. For example, we approximate $K_0(a)\sim\log{2/a}-\gamma\sim-\log a$, where $\gamma$  is the Euler constant. Since we use the boundary condition for $v_z^{(2)}$ at $r=a$ to determine the swimming velocity, we next expand $T_{rz}^{(2)(m,\,n=0)}$ in powers of $a$ to find}
\begin{equation} 
T_{rz}^{(m=n=0)} \sim\frac{ |{\tilde h}|^2}{2} \frac{   \mathrm{De} -
  \mathrm{De}_2 }{1 + \mathrm{De}^2}\frac{\log r}{r \log^2 a}.
\end{equation}
The second
order velocity in the $m=0$, $n=0$ mode is thus
\begin{equation}
v_z^{(2)(m=n=0)} (r)  
= \frac{ |{\tilde h}|^2 \log^2 r}{2 \log^2 a} \frac{\mathrm{De} -
  \mathrm{De}_2}{1 +
  \mathrm{De}^2} - U. 
\end{equation}
\trp{To determine the constant $U$, average the boundary condition (\ref{BC2}) over $t$ and $\phi$ to find}
\begin{widetext}
\begin{eqnarray}
\left. v_z^{(2)(m=n=0)}\right|_{r=a} &=& \left[ -h \cos \phi \partial_r v^{(1)}_z +
  \frac{h}{a} \sin \phi \partial_\phi v^{(1)}_z \right]^{(m=n=0)}_{r=a}\nonumber\\
&=&  | \tilde h|^2/2.
\end{eqnarray}
\end{widetext}
The swimming velocity is therefore
\begin{equation}
U= -\omega q \frac{|\tilde h|^2}{2} \frac{ 1 + \mathrm{De}^2 \eta_s / \eta}{1
  + \mathrm{De}^2},
\label{singlemodeU}
\end{equation}
\trp{where we have reinstated the dimensions and used
  $\mathrm{De}_2=\mathrm{De} \eta_\mathrm{s}/\eta$. The filament swims in the same direction as it would in a Newtonian fluid, opposite the direction of the traveling wave, but with a reduced speed, since $\eta_\mathrm{s}/\eta<1$.}
This result is valid up to leading order in $1/\log a$, and to second
order in $\tilde h$.

\trp{In the appendix we describe the first-order solutions for a longitudinal wave. The same steps just outlined lead to a swimming velocity of} 
\begin{equation}
U_\mathrm{long}= \omega q\frac{|\tilde h|^2}{2} \frac{ 1 + \mathrm{De}^2 \eta_s / \eta}{1
  + \mathrm{De}^2}.
\label{Ulong}
\end{equation}
\trp{For a longitudinal wave, the filament again swims in the same direction as it would in a Newtonian fluid, along the direction of the traveling wave, and with a reduced speed. To see why an object with longitudinal propagating waves swims in the direction of the propagating wave, see reference~\cite{ehlers_etal1996}.}

To apply these results to filaments with generic beating patterns, we
must consider the swimming velocity of a cylinder with a beating
pattern described by a superposition of \trp{transverse or longitudinal} traveling waves,
\begin{equation}
h(s,t) =\sum_{q,\omega}\tilde h_{q,\omega}\mathrm{e}^{\mathrm{i} (q s - \omega t)},\label{hmanymodes}
\end{equation}
written in dimensional form.  
\trp{Since the first-order equations are linear, the first order flow is a superposition of flows corresponding to the individual Fourier components $\tilde h_{q,\omega}$.}
As above, the swimming
velocity is affected only by $T_{rz}^{(m=n=0)}$ and
$G_{rz}^{(m=n=0)}$.  
Due to the averaging
over $t$ involved in obtaining the $m=0$ mode of $T_{rz}$, the first
order flows with different wavelengths do not interact, \trp{yielding} 
\begin{equation}
T_{rz}^{(m=n=0)} = \sum_{q,\omega} T_{rz}^{(m=n=0)}(q,\omega),
\end{equation}
where $T_{rz}^{(m=n=0)}(q,\omega)$ is $T_{rz}^{(m=n=0)}$
calculated for a single traveling wave.  This separation of modes \trp{holds for} 
the swimming velocity, and we obtain the
dimensional 
result
\begin{equation}
U = -\sum_{q,\omega}  \frac{ 1 + (\omega \lambda)^2 \eta_s / \eta}{1
  + (\omega \lambda)^2} q \omega |\tilde h_{q,\omega}|^2.
\label{VEswimspeed}
\end{equation}
\trp{Note that the factor of two difference between Eqs.~(\ref{singlemodeU}) and (\ref{VEswimspeed}) arises since the single wave $\cos(s-t)$ has $\tilde h=1$ in Eq.~(\ref{transverseh}), and $\tilde h_{11}=\tilde h_{-1-1}=1/2$ in Eq.~(\ref{hmanymodes}).}
Finally, we note that
for three-dimensional \trp{transverse} motion,
\begin{equation}
h(s,t) = \sum_{q,\omega}\left[ \hat{ \mathbf{e}}_x
\tilde h^{x}_{q,\omega} \mathrm{e}^{\mathrm{i} (q x - \omega t)} + \hat{ \mathbf{e}}_y  \tilde h^{y}_{q,\omega} \mathrm{e}^{\mathrm{i} (q x - \omega t)}\right],
\end{equation}
\trp{the two polarizations decouple upon averaging over $\phi$ to obtain the $n=0$ mode, leading to}
\begin{equation}
U = -\sum_{q,\omega}  \frac{ 1 + (\omega \lambda)^2 \eta_s / \eta}{1
  + (\omega \lambda)^2} q \omega \left( |\tilde
  h_{q,\omega}^{x}|^2 + |\tilde
  h_{q,\omega}^{y}|^2 \right).
\label{VEswimspeed3D}
\end{equation}

\section{Ramifications for the scallop theorem}
\label{ScallopThm}


\trp{As mentioned in the introduction, the kinematic reversibility of  Stokes equations leads to the scallop theorem for viscous fluids, which states that reciprocal motion of a body leads to no net translation.}  A
reciprocal motion is one which is time reversible, up to a
reparameterization in time. \trp{In this section, we use the swimming velocity in an Oldroyd B fluid, Eq.~(\ref{VEswimspeed}), to examine how the scallop theorem breaks down in viscoelastic fluids.}
Equation~(\ref{VEswimspeed}) is
only valid to second order in displacements, and
all of our statements have the same limitation.  Due to this limitation, motions which we identify as having zero swimming speed may have nonzero swimming speeds at higher order; nonetheless, the reciprocal motions that
have nonzero swimming speed 
in Eq.~(\ref{VEswimspeed}) serve to
identify the leading order violations of the scallop theorem.


We start with some specific classes of beating motions that 
\trp{satisfy the hypotheses of}
the scallop theorem.  
\trp{In these} examples 
\trp{the} motion \trp{is} confined to a
plane, but the conclusions \trp{can be generalized}
to three dimensional motion.
Consider a strictly time-reversal invariant
beating motion, $h(s,t) = h(s,-t)$, \trp{where $h$ is a transverse deflection, and, without loss of generality, the origin has been chosen as the midpoint of the period}.  In this case, the Fourier
coefficients satisfy $h_{q,\omega} = h_{q,-\omega}$,  
and
therefore the sum \trp{in Eq.~(\ref{VEswimspeed})} vanishes.  
\trp{Note that} the same argument can be
made for space reflection symmetric beating patterns as well.  Next,
consider a beating motion which has effectively only one degree of
freedom: 
$h(s,t)
= f(t) g(s)$.  In this case, the Fourier coefficients are also
factorizable, so $h_{q,\omega} = f(\omega) g(q)$.   Since both $f(t)$
and $g(s)$ are real functions, $f(\omega) = f^*(-\omega)$ and $g(q)
=g^*(-q)$, and therefore $|h_{q,\omega}|^2 = |h_{q,-\omega}|^2$.  As a
result, the summand in Eq.~(\ref{VEswimspeed}) is odd in $\omega$, and
the swimming speed vanishes. \trp{More generally, any motion with $|h_{q, \omega}| = |h_{q,-\omega}|$, which corresponds to a superposition of standing waves, leads to no net translation.}


\trp{These} examples demonstrate how time-reversible motions lead to
zero swimming speed \trp{at leading order in an Oldroyd B fluid.  In a Newtonian fluid at zero Reynolds number, more general reciprocal motions, such as a fast power stroke followed by a slow return stroke, also lead to zero swimming speed. To see how this conclusion does not apply to the Oldroyd B fluid, it is convenient to write the swimming speed $U$ in the time domain and in real space,}
\begin{equation}
U=\int_0^T\frac{\mathrm{d}t}{T}\int_0^L\frac{\mathrm{d}s}{L}\int^t_{-\infty}
\mathrm{d}t'
\frac{\partial h}{\partial s}(s,t)M(t,t')\frac{\partial h}{\partial t'}(s,t'),\label{timedomain}
\end{equation}
\trp{where the memory kernel $M(t,t')$ is the operator}
\begin{equation}
\frac{1}{\lambda}\mathrm{e}^{-(t-t')/\lambda }\left(1-\frac{\lambda^2\eta_\mathrm{s}}{\eta}\frac{\partial^2}{\partial t'^2}\right).
\end{equation}
\trp{Any reciprocal motion $h(s,t)$ with period $T$ may mapped to a strictly time-reversal invariant motion by a reparameterization $t_1=F(t)$, where $F$ is monotonic, $F(0)=0$, and $F(T)=T$. For convenience we have chosen the origin as the beginning of the period.
In the Newtonian case,  $\lambda=0$, and the kernel in Eq.~(\ref{timedomain}) becomes a Dirac delta function, $M(t,t')=\delta(t^--t')$, with the interpretation that the singularity of the delta function lies within the region of integration. Inspection of the resulting expression for $U$ reveals that it is reparameterization invariant, and we conclude that our formula for the swimming speed obeys the scallop theorem for general reciprocal motions in a Newtonian liquid at zero Reynolds number. However, when $\lambda\neq0$, the kernel $M(t,t')$ spoils reparameterization invariance, and the speed need not vanish. The physical interpretation is that the distance covered per period in a non-Newtonian fluid depends on the rate of motion as well as the sequence of shapes. }

\trp{For} example, consider 
a helical wave \trp{which propagates} 
forward by one wavelength in the first
third of the time period, and then \trp{backwards by one wavelength} 
in the final
two-thirds of the period:
\begin{widetext}
\begin{equation}
\mathbf{h}(s,t) = \left\lbrace \begin{array}{ll}
 b \cos(s
  -3 t) \hat{\bf {e}}_x + b  \sin(s - 3t) \hat{\bf {e}}_y   
  &0<t<2\pi/3\\
 -b \cos( s+ 3 t/2) \hat{\bf {e}}_x -  b \sin( s + 3t/2) \hat{\bf {e}}_y  & 2\pi/3 < t<2\pi\end{array}\right.\label{waveform}
\end{equation}
\end{widetext}
For this motion,
the Fourier \trp{amplitudes} 
for $\omega=3$ are $h_{1,3}^x=h_{-1,-3}^x=\mathrm{i}h_{1,3}^y=-\mathrm{i}h_{-1,-3}^y=b/6$. For $\omega\neq3$, we have $h_{-1,-\omega}^x=\left(h_{1,\omega}^x\right)^*$, $h_{1,\omega}^y=-\mathrm{i}h_{1,\omega}^x$, and $h_{-1,-\omega}^y=\left(h_{1,\omega}^y\right)^*$, where
\begin{equation}
h_{1,\omega}^x=\frac{9b}{4\pi}\frac{\exp\left(-\mathrm{i}\pi\omega/3\right)\sin\left(\pi\omega/3\right)}{(\omega-3)(\omega+3/2)}.
\end{equation}
All other $h_{q,\omega}$ are zero.
Note that these coefficients do not obey the criterion $|h_{q,\omega}|
= |h_{q, -\omega}|$.  
\trp{Inserting these coefficients into Eq.~(\ref{VEswimspeed3D}) and using contour integration to evaluate the resulting sums~\cite{CarrierKrookPearson1966} yields the dimensionless speed}
\begin{widetext}
\begin{equation}
U = -\left(1- \frac{\eta_s}{\eta}\right)\frac{27 \mathrm{De}^2 b^2 }{\pi}\frac{\left[- \pi(4+ 45
  \mathrm{De}^2 + 81 \mathrm{De}^4) + \frac{ 18 \mathrm{De} (2 + 9
  \mathrm{De}^2 )\sinh \frac{2 \pi}{3 \mathrm{De}}}{ 1 + 2 \cosh{\frac{2
  \pi}{3 \mathrm{De}} } }\right]}{(4 + 45 \mathrm{De}^2 + 81
  \mathrm{De}^4)^2},\label{Ufastslow}
\end{equation}
\end{widetext}
where $\mathrm{De} = 2 \pi\lambda/T$ \trp{and $T$ is the dimensional period of the motion.} 

\begin{figure}
\includegraphics[width=9.4cm]{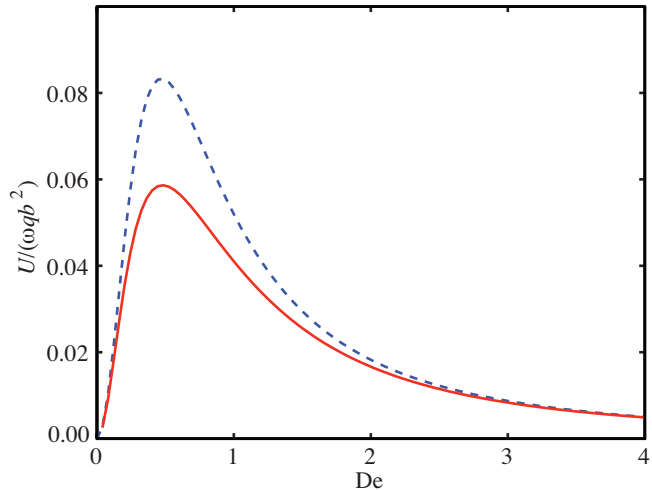}
\caption{(Color online) The swimming velocity of the reciprocal motion of Eq.~(\ref{waveform}) versus $\mathrm{De} = 2 \pi
  \lambda/T$, for $\eta_s/\eta = 0.75$.
The (red) solid curve is the exact second-order result of Eq.~(\ref{Ufastslow}); the (blue) dashed curve corresponds to the estimate of Eq.~(\ref{estimated}).}
%
\label{scallopfig}
\end{figure}

\trp{The swimming speed vanishes in the Newtonian limits of no
  polymer, $\eta_\mathrm{s}=\eta$, or zero relaxation time,
  $\mathrm{De}=0$.}  For these cases, the swimmer moves in the negative direction during the first
third of the period, and in the positive direction by the exact same amount during the
last two thirds of the period.  \trp{In the viscoelastic case, the
  swimmer moves more slowly during the first third of the period when
  the wave frequency is high. Thus, the scallop theorem is rendered
  inapplicable and the net motion of the swimmer is in the positive
  (forward) direction. The formula~(\ref{Ufastslow}) for the swimming speed is complicated due to memory effects: the speed of the swimmer during each stroke depends on its motion during previous strokes. However, we can make a simple estimate for the swimming speed by disregarding these memory effects and using Eq.~(\ref{singlemodeU}) to estimate the speed during each stroke of the waveform of Eq.~(\ref{waveform}):}
\begin{equation}
U_{\mathrm{est}} =  b^2  \left[\frac{1+(9/4)\mathrm{De}^2(\eta_s/\eta)}
{1 + (9/4)\mathrm{De}^2} -
 \frac{1+9\mathrm{De}^2(\eta_s/\eta)}{1 + 9\mathrm{De}^2}\right].\label{estimated}
\end{equation}
Figure~\ref{scallopfig} shows the \trp{exact second-order~(\ref{Ufastslow}) and estimated}~(\ref{estimated}) swimming speeds 
\trp{for the waveform~(\ref{waveform}), with dimensional wavelength $2\pi/q$.} 

Helical waves 
provide a
convenient experimental system to explore the fate of the scallop
theorem in viscoelastic fluids, since the backward and forward
traveling waves could be easily actuated by rotating a rigid helical
filament \trp{wrapped in a flexible rubber sleeve~\cite{gi_taylor1952}}. 
For a cylinder of radius $a=1$\,cm
deformed into 
\trp{a helix} with pitch $2\pi/q=12$\,cm and helical radius $h=0.5$\,cm
rotating with a period $T=1.5$\,s,
the swimming velocity of a helix in a Newtonian fluid is $U=\omega q h^2\sim 0.5$\,cm/s, which corresponds to unity on the vertical axis of Fig.~\ref{scallopfig}.  Note that these values are chosen \trp{to make} 
$aq \approx 0.5$, $bq\approx
0.25$, and $h/a \approx 0.5$.
The time constant needed to place the swimming velocity around
the maximum value would be on the order of 0.1\,s, which is within the
range of time constants \trp{found} 
in Boger fluids
\cite{BinningtonBoger1986}.    Also assuming that 
$1- \eta_s/\eta = \eta_p/\eta_s \approx 0.25$, again typical for
Boger fluids \cite{BinningtonBoger1986, Phan-Thien1984}, the swimming
velocity for our example time-varying reciprocal motion is $0.1$--$0.5$\,mm/s.  Note that larger swimming speeds might be obtained by
using fluids with smaller time constant $\lambda$, but care must be
taken to remain in the low Reynolds number regime.  For the parameters
we used above, the Reynolds number 
is less than 0.01 for a Boger fluid with viscosity in the range 0.2--0.5\,Pa-s.  

\section{Conclusion}\label{discuss}

In this paper we have calculated the swimming velocity for an infinite
cylinder undergoing arbitrary periodic beating motion in a nonlinearly
viscoelastic Oldroyd-B fluid.  We have assumed that the amplitude of
deflection 
is small
compared to the wavelength of deformations and compared to the radius
of the cylinder.  We find that in general, nonlinear viscoelastic
corrections decrease the swimming velocity relative to the Newtonian
case.  It is interesting to note that the decrease in swimming
velocity depends on the nonlinearity:  in a linear Maxwell fluid, even
though the time-reversal invariance of the constitutive relation is
manifestly broken by memory effects through the time derivative of the
stress, for a given swimming motion, the flows are exactly the same as
the Newtonian case (although the stresses are modified)~\cite{FulfordKatzPowell1998}.  Nonlinearity in the
constitutive relation is required to rectify the changes in the stress
into changes of flows.    
\trp{We have used our calculation to examine the breakdown of the
  scallop theorem in a nonlinearly viscoelastic fluid. Within our
  assumptions of small amplitude deflections of an infinitely long filament,
  time-reversal-invariant motions lead
  to no net translation. We further showed that net motion in a
  nonlinearly viscoelastic fluid is possible for a more general
  reciprocal motion, in which the backwards and forwards strokes occur
  at different rates.  These predictions could be readily tested with
  simple table-top experiments. The nonlinearity of our constitutive
  relation necessarily limited our analytical results to the case of
  small deflection. It would be of great value to develop a proper
  local resistance theory for slender rods in a nonlinearly
  viscoelastic fluid.}

In this paper we have worked in the frequency
domain, building arbitrary motions through superposition.  
This allows us to easily treat time-periodic
motions, but does not lend itself to a physical understanding of
motions that suddenly stop or start, which require a complicated
superposition of many frequencies.  In numerical studies of
peristaltic flows, Teran, Fauci, and
Shelley~\cite{TeranFauciShelley2008} have explored the build-up of
stresses from an initial fluid stress given by an isotropic pressure.  
In this context, time-periodic flows correspond to asymptotic
long-time behavior, which may take a long time to develop in numerical
simulations.  The numerical studies provide a complementary viewpoint
well-suited to revealing transient behavior in flows.

This work was supported in part by National Science Foundation Grants Nos. NIRT-0404031 (TRP) and DMS-0615919 (TRP), and NIH grant No. R01
GM072004 (CWW).
TRP thanks the Aspen Center for Physics, where some of this work was done. We thank Allen Bower and Eric Lauga for helpful discussions.

\appendix

\section{Forces on a cylinder with prescribed beating pattern.}
\label{forces1st}

The force per unit length acting on the cylinder is found by
integrating the stress tensor \trp{around the circumference,}
\begin{equation}
{\bf  f} (s) = a \int {\bf \hat n}(\psi,z)  \cdot {\bm \sigma}(
r_\mathrm{b},\phi_\mathrm{b},z_\mathrm{b}) \mathrm{d}\psi, \label{feqn}
\end{equation}
\trp{where the dependence of the cylindrical coordinates $\{r_\mathrm{b}, \phi_\mathrm{b}, z_\mathrm{b}\}$ on $\psi$ and $s$ depends on the deformation. First consider the force per unit length $\mathbf{f}_\perp$ for a transverse wave with polarization in the $x$-direction only, Eq.~(\ref{transverseh}). Mirror symmetry about the $xz$-plane implies that $f_{\perp y}=0$. }
To lowest order, 
$\hat {\mathbf n} = \hat{\mathbf r}$, 
$r_\mathrm{b}=a$, $\phi_\mathrm{b}=\psi$, and $z_\mathrm{b}=s$ [see Eq.~(\ref{boundaryposition})].
 Upon integration
over $\psi$, the force in the $\hat \mathbf z$-direction arising from $\sigma^{(1)}_{r z} = \tau_{r z}^{(1)}$ vanishes, and
the force per unit length 
arising from $\sigma^{(1)}_{rr} = - p^{(1)} +
  \tau_{rr}^{(1)}$ and $\sigma^{(1)}_{r \phi} = \tau_{r \phi}^{(1)}$ is
\begin{equation}
{\bf f}^{(1)}_\perp = \mathrm{Re} \left\lbrace \frac{4\pi}{\log a} \frac{1
  - \im \mathrm{De}_2}{1
  - \im \mathrm{De}} \left(-\im
  \tilde h \mathrm{e}^{\im (s-t)}\right)\right\rbrace \hat {\bf x} 
\label{firstforce}
\end{equation}
 This force is 
 accurate
 to lowest order in $1/\log(a)$.  

Now 
\trp{consider} longitudinal traveling waves:
\begin{equation}
{\bf r}(s,\psi) =  \hat{\bf x}a \cos \psi  + \hat {\bf y}a \sin\psi  + \hat {\bf z}\mathrm{Re}[\tilde h \mathrm{e}^{\im (qs - \omega t)} + s] . 
\end{equation}
Using cylindrical coordinates, 
the position of the surface can also be described by
$(r_\mathrm{b},\phi_\mathrm{b},z_\mathrm{b})$.   To first order in $h$, 
\begin{eqnarray}
r_\mathrm{b} &=& a \nonumber \\
\phi_\mathrm{b} &=& \psi \nonumber \\
z_\mathrm{b} &=& s + h \label{boundarypositionLongitudinal}
\end{eqnarray}
\trp{As in the transverse case,} the no-slip boundary conditions at the filament surface are \trp{found by expanding the argument of the flow velocity at the surface to first order,}
\begin{equation}
\left[\mathbf{v} +h\hat{\bf z}\cdot\nabla\mathbf{v}\right]_{a,\psi,s,t}=\dot h\hat{\bf z}.
\end{equation}
The nondimensional first and second order flow equations are the same as in the
transverse case, but the boundary condition is different.
\trp{At first order,} $\tilde v_z(a,\psi) =
-\im \tilde h \hat {\mathbf{z}}$.  
The 
first order
velocity field and \trp{pressure} 
are \trp{shown} in Table \ref{firstorderlongitudinal}.
\trp{Using these solutions in general expression for the force per unit length, Eq.~(\ref{feqn}), we find that the} first order force for a longitudinal traveling wave is 
\begin{equation}
{\bf f}^{(1)}_\parallel = \mathrm{Re} \left\lbrace \frac{2 \pi}{\log a} \frac{1
  - \im \mathrm{De}_2}{1
  - \im \mathrm{De}} \left[-\im
  \tilde h \mathrm{e}^{\im (s-t)}\right]\right\rbrace \hat {\bf z} 
\label{firstforcelongitudinal}
\end{equation}

\begin{table*}
\begin{tabular}{|c||c|}
\hline \\
$\tilde p $& $\tilde h  K_0(r)$\\[1ex]
 $\tilde v_r $& $\tilde h  \left[\alpha A r K_0(r)/2+BK_1(r)\right] $\\[1ex]
$\tilde v_\phi $& $0$\\[1ex]
$\tilde v_z $& $-\mathrm{i} \tilde h \left[(B-\alpha A) K_0(r)+\alpha a A r K_1(r)/2\right] $\\[1ex]
$\alpha A$ & $-2K_1(a)/\left\{a\left[K_0(a)^2-K_1(a)^2\right]+2K_0(a)K_1(a)\right\}$\\[1ex]
$ B$&$aK_0(a)/\left\{a\left[K_0(a)^2-K_1(a)^2\right]+2K_0(a)K_1(a)\right\}$\\
\hline 
\end{tabular}
\caption{ \trp{Pressure and velocity field for a cylinder with a longitudinal traveling wave; $\alpha=(1-\mathrm{i De})/(1-\mathrm{i De}_2)$.}
 }
\label{firstorderlongitudinal}
\end{table*}

As explained before, the first order force is sufficient for
calculating the lowest order changes to swimming properties that
result from shape changes of beating patterns in viscoelastic media.  
To apply Eq.~(\ref{firstforce}), we interpret it in a manner
consistent with resistive force theory~\cite{gray_hancock1955}.   \trp{Since} the quantity $-\im
  \tilde h \exp(\im s-\im t)$ is the transverse velocity of the cylinder, 
  the force per unit length has a complex transverse \trp{resistance} coefficient of 
$-(4\pi/\log a) (1 - \im \mathrm{De}_2)/(1 - \im\mathrm{De})$
for motion occurring with dimensional space- and time-dependence
$\exp(\im q s- \im \omega t)$.  This 
\trp{resistance} coefficient depends on $q$ through the
dimensionless parameter $a$.  As is commonly done in resistive force
theory, to treat general motion involving a superposition of traveling waves
with multiple wavelengths, we replace the slow logarithmic dependence
on $q$ with a constant involving a typical wavevector $\langle q \rangle$.
The error introduced in this process is of order $(q - \bar
  q)/(\bar{q} \log \bar q)$.
After this replacement we obtain a \trp{resistance} 
coefficient, 
in dimensional form, of
\begin{eqnarray}
\zeta_{\perp}^\mathrm{Oldroyd}& = &-\zeta_\perp 
\frac{1 - \im\mathrm{De}_2}{1 - \im \mathrm{De}}\\
  \zeta_\perp&=&-\frac{4\pi\eta}{\log \langle q \rangle a}.
\end{eqnarray}
The parallel \trp{resistance} 
coefficient is obtained by a similar analysis for
the longitudinal wave, and is given by $\zeta^\mathrm{Oldroyd}_\parallel = \zeta^\mathrm{Oldroyd}_\perp/2$ and $\zeta_\parallel = \zeta_\perp/2$.
\trp{Generalization to motion with multiple frequencies or helical waves may be accomplished by linear superposition.}
  This choice of $\zeta_\perp$ and
  $\zeta_\parallel$  is consistent with
  those previously introduced for the Newtonian resistive
  force theory. 

With this choice of resistance coefficients, 
the \trp{first-order} force
per unit length obeys 
\begin{equation}
\mathbf{f}
+\lambda
 \dot\mathbf{f}
 =\mathbf{f}_\mathrm{vis} +
 \frac{\eta_s}{\eta} \lambda \dot \mathbf{f}_\mathrm{vis}, 
 \label{rftOldroyd}
\end{equation}
\trp{where $\mathbf{f}_\mathrm{vis}$ is the force per unit length from Newtonian resistive force theory,}
\begin{equation}
\mathbf{f}_\mathrm{vis}(s,t)=\zeta_\perp\mathbf{v}_\perp(s,t)+\zeta_{||}\mathbf{v}_{||}(s,t).\label{RFTNewt}
\end{equation}
\trp{and $\mathbf{v}_\perp(s,t)$ and $\mathbf{v}_{||}(s,t)$ are the perpendicular and parallel components of the flow velocity \textit{relative to the rod} at the point $s$ on the rod. The velocity $\mathbf{v}$ must be interpreted as a relative velocity to maintain Galilean invariance.} In the limit of \trp{zero} 
solvent viscosity, $\eta_s = 0$, 
\trp{these equations are equivalent to} 
the resistive force theory proposed by Fulford,
Katz, and Powell for the linear Maxwell model~\cite{FulfordKatzPowell1998}.  

\trp{We can now see why the resistive force theory (\ref{rftOldroyd}) leads to no change in swimming speed relative to Newtonian resistive force theory. First note that in resistive force theory, the swimming velocity of a body is determined by demanding that the total force $\mathbf{F}=\int\mathrm{d}s\,\mathbf{f}$ on the body vanishes. Integrating Eq.~(\ref{rftOldroyd}) over arclength $s$ and then averaging over a period leads to the condition $\int\mathrm{d}s\,\mathbf{f}_\mathrm{vis}=0$, the same condition for determining the swimming velocity in the Newtonian case.}

\trp{Although  Eq.~(\ref{rftOldroyd}) gives internally consistent predictions for the swimming speed of a filament with a traveling wave, it is nevertheless flawed. To see why, recall our motivation of the nonlinear constitutive relation for a fluid with fading memory in Sec.~\ref{models}. There we saw that without the nonlinear terms of the upper-convective derivative, the Maxwell constitutive relation for an element is sensitive to the local rate of rotation of that element. To illustrate this effect, consider shear flow in a channel on a slowly rotating table. A short calculation shows that the zero-frequency shear viscosity depends on the rate of rotation of the table, an unphysical result if the rate of rotation is low enough that the fictitious forces acting on the polymers in solution are small compared to the viscous forces~\cite{BirdArmstrongHassager1987}.}

\trp{Likewise, we may consider a slender filament on a table rotating with speed $\Omega$, and ask if Eq.~(\ref{rftOldroyd}) predicts forces that are independent of $\Omega$. Set $\eta_\mathrm{s}=0$ for convenience. Both the fluid and the filament are rotating with the same speed. Suppose further that the filament is moved with steady velocity $\mathbf{v}$ relative to the fluid. Unless $\Omega$ is enormously large, we expect the force per unit length $\mathbf{f}$ to eventually relax to the viscous force per unit length~(\ref{RFTNewt}). However, writing Eq.~(\ref{rftOldroyd}) in terms the perpendicular and parallel components, and solving for the force in the steady state yields}
\begin{eqnarray}
f_{||}&=&\frac{\zeta_{||}v_{||}+\lambda\Omega\zeta_\perp v_\perp}{1+\lambda^2\Omega^2}\\
f_\perp&=&\frac{\zeta_\perp v_\perp-\lambda\Omega\zeta_{||} v_{||}}{1+\lambda^2\Omega^2}.
\end{eqnarray}
\trp{Thus, we see explicitly that the resistive force theory for the Maxwell constitutive relation predicts spurious dependence on the rotation speed $\Omega$. Equation~(\ref{rftOldroyd}) is only truly reliable when working to linear order in deflection.}

\end{document}